\documentclass{emulateapj}
\usepackage{mathrsfs}
\usepackage{CJK}

\usepackage{amsmath}

\usepackage{graphics}

\newcommand{\bd}{\begin{displaymath}}
\newcommand{\ed}{\end{displaymath}}

\slugcomment{Received 2013 October 29; accepted 2014 May 27;
published 2014 June 23}

\shorttitle{Episodic jet power extracted from spinning black holes
in GRB}

\begin{document}

\title{Episodic jet power extracted from a spinning black hole
surrounded by a neutrino-dominated accretion flow in gamma-ray
bursts}

\author{Xinwu Cao\altaffilmark{1,2}, En-Wei Liang\altaffilmark{3}, and Ye-Fei Yuan\altaffilmark{4}}
\altaffiltext{1}{Key Laboratory for Research in Galaxies and
Cosmology, Shanghai Astronomical Observatory, Chinese Academy of
Sciences, 80 Nandan Road, Shanghai, 200030, China; cxw@shao.ac.cn}
\altaffiltext{2}{Department of Astronomy and Institute of
Theoretical Physics and Astrophysics, Xiamen University, Xiamen,
Fujian, 361005, China} \altaffiltext{3}{Department of Physics and
GXU-NAOC Center for Astrophysics and Space Sciences, Guangxi
University, Nanning, 530004, China; lew@gxu.edu.cn}
\altaffiltext{4}{Department of Astronomy, University of Sciences and
Technology of China, Hefei, Anhui, 230026, China;
yfyuan@ustc.edu.cn}

\begin{abstract}

It was suggested that the relativistic jets in gamma-ray bursts
(GRBs) are powered via the Blandford-Znajek (BZ) mechanism or the
annihilation of neutrinos and anti-neutrinos from a neutrino
cooling-dominated accretion flow (NDAF). The advection and diffusion
of the large-scale magnetic field of a NDAF is calculated, and the
external magnetic field is found to be dragged inward efficiently by
the accretion flow for a typical magnetic Prandtl number
$\mathscr{P}_{\rm m}=\eta/\nu\sim 1$. The maximal BZ jet power can
be $\sim 10^{53}-10^{54}$~erg~s$^{-1}$ for an extreme Kerr black
hole, if an external magnetic field with $10^{14}$~Gauss is advected
by the NDAF. This is roughly consistent with the field strength of
the disk formed after a tidal disrupted magnetar. The accretion flow
near the black hole horizon is arrested by the magnetic field if the
accretion rate is below than a critical value for a given external
field. The arrested accretion flow fails to drag the field inward
and the field strength decays, and then the accretion re-starts,
which leads to oscillating accretion. The typical timescale of such
episodic accretion is in an order of one second. This can
qualitatively explain the observed oscillation in the soft extend
emission of short-type GRBs.
\end{abstract}



\section{Introduction}
It is believed that gamma-ray bursts (GRBs) are powered by the
accretion disks surrounding stellar-mass black holes
\citep*[][]{1999ApJ...518..356P,2001ApJ...557..949N}. The observed
high luminosity of GRBs requires a very large mass accretion rate,
$\dot{M}\sim 0.01-10~M_{\odot}{\rm s}^{-1}$, and therefore the
accretion disks are massive with mass of $0.1-1~M_\odot$. The gas in
the accretion disks is so dense that the most photons emitted in the
disks cannot escape the systems, which leads to high temperature of
the disks, $\ga 10^{10}$~K. Thus, the neutrino cooling becomes
dominant over the photon radiation, and therefore the accretion
disks are named as ``neutrino-dominated accretion flows"
\citep*[NDAF,][]{1999ApJ...518..356P,2001ApJ...557..949N,2002ApJ...579..706D,2002ApJ...577..311K,2005ApJ...629..341K,2006ApJ...643L..87G,2007ApJ...657..383C,2007ApJ...662.1156K,2007ApJ...661.1025L,2012ApJ...759...82P}
.

The widely discussed jet formation mechanisms for GRBs are the
Blandford-Znajek (BZ) mechanism or the annihilation of neutrinos and
anti-neutrinos from an NDAF
\citep*[e.g.,][]{1999ApJ...518..356P,2000ApJ...533L.115L,2001ApJ...557..949N,2002ApJ...579..706D,2002ApJ...577..311K,2002MNRAS.335..655W,2005ApJ...630L...5M,2005ApJ...632..421L,2006ApJ...643L..87G,2007ApJ...657..383C,2008ApJ...687..433J,2009ApJ...700.1970L,2012ApJ...760...63L}.
Based on lack of detection of a thermal photosphere component
predicted by the baryonic fireball model in the GRB spectra observed
with the {\em Fermi} mission
\citep*[e.g.,][]{2009Sci...323.1688A,2011ApJ...730..141Z}, it was
proposed that the GRB jet would be highly magnetized (e.g., Zhang \&
Pe¡¯er 2009; Zhang \& Yan 2011; see also Daigne \& Mochkovitch 2002;
Zhang \& M¡äesz¡äaros 2002). More recently,
\citet{2012ApJ...757...56Y} proposed a new model for the jet
formation in GRBs. The episodic, magnetically dominated plasma blobs
are suggested to be ejected from the corona of a hyper accretion
disk, which are responsible for the observed episodic prompt
gamma-ray emission from GRBs, as suggested by
\citet{2011ApJ...726...90Z}.

The most extensively discussed mechanism for powering a magnetized
jet is the BZ mechanism \citep*[][]{1977MNRAS.179..433B}, in which
the energy and angular momentum are extracted from a rotating black
hole with the large scale magnetic field lines. The power of the jet
launched by a rotating black hole surrounded by a NDAF can be
estimated when the strength of the magnetic field near the black
hole horizon is available. \citet{2013ApJ...766...31K} calculated
the BZ power of the jet in GRBs, and found that the typical jet
power can be as high as $\sim10^{53}~{\rm erg~s}^{-1}$ assuming the
poloidal field to be equipartition with the gas pressure in the
inner disk region. In their work, the detailed physics of the large
scale field formation has not been considered.

A NDAF is probably formed after a merger of black hole/neutron
binary
\citep*[e.g.,][]{1989Natur.340..126E,1998A&A...338..535R,2010MNRAS.401.1644B,2012PhRvD..85f4029E,2013ApJ...776...47D,2013PhRvD..87h4006F}
or the gravitational collapse of a massive star
\citep*[e.g.,][]{1993ApJ...405..273W,1998ApJ...494L..45P,2004RvMP...76.1143P,2004IJMPA..19.2385Z,2006ARA&A..44..507W}.
In the merger scenario, a disk is formed to feed the black hole
after a neutron star is tidally disrupted, and the strong magnetic
field may be formed near the black hole with an external field
dragged inward by the NDAF. The NDAF is driven by turbulence, which
also makes the field diffuse outwards. A steady magnetic field
configuration of the disk is available when the advection is
balanced with the diffusion of the field in the disk
\citep{1994MNRAS.267..235L}.

The diffusivity $\eta$ of the magnetic field is closely related to
the viscosity $\nu$. The magnetic diffusivity should be at the same
order of the viscosity in isotropic turbulent fluid, i.e.,
$\mathscr{P}_{\rm m}=\eta/\nu\sim 1$ \citep*[][]{parker1979}. This
is indeed conformed by many different numerical simulations
\citep*[e.g.,][]{2003A&A...411..321Y,2009A&A...504..309L,2009A&A...507...19F,2009ApJ...697.1901G}.
The diffusion/advection balance of the large scale magnetic field in
the accretion disks has been explored in some previous works
\citep*[][]{1994MNRAS.267..235L,2009ApJ...701..885L,2011ApJ...737...94C,2009ApJ...707..428B,2012MNRAS.424.2097G,2013MNRAS.430..822G,2013ApJ...765..149C}.
The advection of the field is found to be efficient for ADAFs
\citep{2011ApJ...737...94C}, due to their relative high radial
velocities \citep*[][]{1994ApJ...428L..13N}, while it is inefficient
for the geometrically thin standard accretion disks
\citep{1994MNRAS.267..235L}, unless the thin disks are dominantly
driven by the outflows rather than the turbulence
\citep{2013ApJ...765..149C}.

In this work, we calculate the magnetic field advection/diffusion in
a NDAF, and then estimate the maximal jet power can be extracted
from a rapidly rotating black hole in GRBs. We describe the model in
Section \ref{model}, and Sections \ref{results} and \ref{discussion}
contain the results and discussion.

\section{Model}\label{model}

We assume a vertical and homogeneous external magnetic field to be
radially advected by the gas in the NDAF. The global solution of a
steady NDAF is available by solving a set of differential equations,
namely, the radial and azimuthal momentum equations, the continuity
equation, and the energy equation. The energy equation is rather
complicated for a NDAF \citep*[see,
e.g.,][]{1999ApJ...518..356P,2001ApJ...557..949N,2002ApJ...579..706D,2002ApJ...577..311K,2005ApJ...629..341K,2006ApJ...643L..87G,2007ApJ...657..383C,2007ApJ...662.1156K,2007ApJ...661.1025L}.
Both the photodisintegration cooling and neutrino cooling are
included, which are functions of mass accretion rate. The magnetic
field advection/diffusion is predominantly determined by the radial
velocity of the accretion flow. Similar to the ADAF case, the
dynamics of the NDAF are also altered in the presence of the
large-scale field, especially in the inner region of the flow, which
is properly considered in our calculations. In the following two
sub-sections, we briefly summarize the calculations of the structure
of a NDAF and its magnetic field configuration.

\subsection{Structure of the NDAF}\label{stru_ndaf}

The dynamics of a steady NDAF with a large-scale magnetic field can
be derived by solving the following differential equations.

The continuity equation is
\begin{equation}
{\frac {d}{dR}}(\rho RHv_R)=0, \label{continuity_1}
\end{equation}
which reduces to $-2\pi R\Sigma v_{R}=\dot{M}$ if the winds from the
NDAF are neglected, $\Sigma=2H\rho$, and $\dot{M}$ is the mass
accretion rate of the NDAF.

The radial momentum equation is
\begin{equation}
v_R{\frac {dv_R}{dR}}=-(\Omega_{\rm K}^2-\Omega^2)R-{\frac
1{\rho}}{\frac {d}{dR}}(\rho c_{\rm s}^2)+{\frac {B_R^{\rm
S}B_z}{2\pi \Sigma}}-{\frac {B_z H}{2\pi\Sigma}}{\frac {\partial
B_z}{\partial R}}, \label{radial_1}
\end{equation}
where the pseudo-Newtonian potential is used to simulate the black
hole gravity \citep{1980A&A....88...23P}, $B_R^{\rm S}$ is the
radial magnetic field strength at the disk surface $z=\pm H$. The
Keplerian angular velocity is $\Omega_{\rm K}^2(R)=GM_{\rm
bh}/(R-R_{\rm S})^2 R$, and $R_{\rm S}=2GM_{\rm bh}/c^2$.

As suggested by \citet{1991ApJ...376..214B}, the radial magnetic
fields are sheared into azimuthal fields by the differential
rotation of the accretion flow, which triggers magnetorotational
instability (MRI) and the MRI-driven turbulence
\citep{1998RvMP...70....1B}. In this case, the viscosity can still
be approximated as the conventional $\alpha$-viscosity, though the
values of $\alpha$ are somewhat diverse, as shown in the numerical
simulations. It is found that the final results of the field
advection in the disks are almost independent of the value of
$\alpha$ \citep*[see][for the
discussion]{1994MNRAS.267..235L,2011ApJ...737...94C}. The angular
momentum equation reads
\begin{equation}
{\frac {d\Omega}{dR}}={\frac {v_R(j-j_{\rm in})}{\nu R^2}},
\label{azimuthal_1}
\end{equation}
where $j$ is the specific angular momentum of the accretion flow,
and $j_{\rm in}$ is the specific angular momentum of the gas
accreted by the black hole. The conventional $\alpha$-viscosity
$\nu=\alpha c_{\rm s}H$ is adopted in our calculations.

The energy equation is
\begin{equation}
\Sigma v_R T{\frac {ds}{dR}}=Q^{+}-Q^{-}, \label{energy_1}
\end{equation}
{where $T$ and $s$ are the temperature and entropy of the gas, the
cooling rate $Q^-=Q_{\rm ph}^-+Q_\nu^-$ ($Q_{\rm ph}^-$ and
$Q_\nu^-$ are the cooling rates due to photodisintegration and
neutrino processes, respectively). Compared with these two terms,
the radiation cooling is always much smaller
\citep*[e.g.,][]{2013arXiv1306.0655X}, and it is therefore neglected
in our calculations. The neutrino cooling rate is
\begin{equation}
Q_\nu^-=\sum_{i}{\frac {2(7/8)\sigma
T^4}{(3/4)[\tau_{\nu_i}/2+1/\sqrt{3}+1/(3\tau_{{\rm a},\nu_i})]}},
 \label{qminus_nu_1}
\end{equation}
where $\tau_{\nu_i}=\tau_{{\rm s},\nu_i}+\tau_{{\rm a},\nu_i}$ is
the sum of the optical depth due to scattering and absorption for
each neutrino flavor ``$i$" in the vertical direction. They are
functions of $\rho$, $T$, and $H$ \citep*[see][for the
details]{2002ApJ...579..706D}. The photodisintegration cooling is
\citep*[][]{2002ApJ...579..706D}
\begin{equation}
Q_{\rm ph}^-\simeq 10^{29}\rho_{10}v_{R}{\frac {dX_{\rm
nuc}}{dR}}2H~~ {\rm ergs~cm^{-2}~s^{-1}}, \label{qminus_ph_1}
\end{equation}
where $\rho_{10}=\rho/10^{10}~{\rm g~cm}^{-3}$, and the mass
fraction $X_{\rm nuc}$ of free nucleons is approximately given by
\begin{equation}
X_{\rm nuc}\simeq
34.8\rho_{11}^{-3/4}T_{11}^{9/8}\exp(-0.61/T_{11}). \label{x_nuc_1}
\end{equation}
Here $T_{11}=T/10^{11}~{\rm K}$. }


A set of equations described above are closed with an additional
equation, i.e., the equation of state
\citep*[][]{1999ApJ...518..356P},
\begin{equation}
p={\frac {1+3X_{\rm nuc}}{4}}\rho RT+{\frac
{11}{12}}aT^4+K\left({\frac {\rho}{\mu_{\rm e}}}\right)^{4/3},
\label{state_1}
\end{equation}
where {$K=(2\pi hc/3)(3/8\pi m_{\rm n})^{4/3}=1.24\times 10^{15}$
($m_{\rm n}$ is the nucleon mass), and $\mu_{\rm e}$ is the mass per
electron ($\mu_{\rm e}=2$ is adopted as the previous works).} The
three terms in Equation (\ref{state_1}) are contributed by gas
pressure, pressure due to radiation and relativistic
electron-positron pairs, and degeneracy pressure from relativistic
electrons, respectively. The corresponding expression for the
internal energy is
\begin{equation}
u={\frac {1}{\gamma_{\rm gas}-1}}{\frac {1+3X_{\rm nuc}}{4}}
RT+{\frac {11}{4}}{\frac {aT^4}{\rho}}+3K\left({\frac
{\rho}{\mu_{\rm e}}}\right)^{1/3}, \label{u_1}
\end{equation}
where $\gamma_{\rm gas}$ is the ratio of specific heats for baryonic
gas. The baryonic gas in NDAFs is non-relativistic, $\gamma_{\rm
gas}\simeq 5/3$. The internal energy can be related to pressure as
\begin{equation}
u={\frac {1}{\gamma-1}}{\frac {p}{\rho}}, \label{u_2}
\end{equation}
for two extreme cases, i.e., baryonic gas dominant ($\gamma\simeq
5/3$) or the pressure from relativistic particles and radiation
dominant ($\gamma\simeq 4/3$). It is found that the baryonic gas
pressure is dominant in a NDAF for a wide mass accretion rate range
\citep{2013ApJ...766...31K}. For simplicity, we adopt $\gamma=5/3$
in calculating the structure of the NDAF. Thus, Equation
(\ref{energy_1}) can be re-written as
\begin{displaymath}
{\frac {2}{(\gamma-1)c_{\rm s}}} {\frac {dc_{\rm s}}{dR}}=-{\frac
{1}{R}}-{\frac {1}{H}}{\frac {dH}{dR}}-{\frac {1}{v_R}}{\frac
{dv_R}{dR}}~~~~~~~~~~~~~~~~~~~~~~~~~~~
\end{displaymath}
\begin{equation}
+{\frac {v_R}{\alpha c_{\rm s}^3R^2H}}(j-j_{\rm in})^2+{\frac {2\pi
RQ^-}{\dot{M}c_{\rm s}^2}}, \label{energy_2}
\end{equation}
in which the continuity equation (\ref{continuity_1}) is
substituted. The final results on the advection of the fields in the
NDAFs are found to be insensitive to the value of $\gamma$ adopted.

The accretion flow is vertically compressed by the strong
large-scale magnetic fields
\citep{2002A&A...385..289C,2011ApJ...737...94C}. {Assuming the
pressure gradient to be in equilibrium with gravity in the vertical
direction of the accretion flow, the vertical structure of the disk
can be described by
\begin{equation}
{\frac {dp(z)}{dz}}=-\rho(z)\Omega_{\rm K}^2z-{\frac
{B_{R}}{4\pi}}{\frac {\partial B_R}{\partial z}}+{\frac
{B_{R}}{4\pi}}{\frac {\partial B_z}{\partial R}}, \label{vertical_1}
\end{equation}
if the magnetic field line shape in the disk is known
\citep{2011ApJ...737...94C}. In principle, the shape of the field
line is available by solving the radial and vertical momentum
equations of the disk with suitable boundary conditions
\citep*[see][for the detailed discussion]{2002A&A...385..289C},
which is too complicated for our present work. To avoid this
complexity, we only consider the simplest case assuming the sound
speed $c_{\rm s}$ to be constant in the vertical direction of the
accretion flow, i.e., $dp/dz=c_{\rm s}^2 d\rho/dz$. In this case, an
approximate analytical expression for the field line shape:
\begin{equation}
R-R_{\rm i}={\frac {H}{\kappa_0\eta_{\rm i}^2}}(1-\eta_{\rm
i}^2+\eta_{\rm i}^2z^2H^{-2})^{1/2}-{\frac {H}{\kappa_0\eta_{\rm
i}^2}}(1-\eta_{\rm i}^2)^{1/2}, \label{b_shape_1}
\end{equation}
is proposed, where $H$ is the scale height of the disk [defined as
$\rho(H)=\rho(0)\exp(-1/2)$], the field line inclination
$\kappa_0=B_z/B_R^{\rm S}$ at $z=H$, and $\eta_{\rm i}=\tanh(1)$
\citep{2002A&A...385..289C}. The main features of the
Kippenhahn-Schl\"{u}ter model \citep{1957ZA.....43...36K} can be
well reproduced by this expression either for weak or strong field
cases. The disk scale height $H$ can therefore be derived
numerically with the magnetic field line shape (\ref{b_shape_1}). We
adopt a fitting formula suggested by \citet{2011ApJ...737...94C} to
calculate the disk scale height in this work,
\begin{equation}
{\frac {H}{R}}={\frac {1}{2}}\left ({\frac {4c_{\rm
s}^2}{R^2\Omega_{\rm K}^2}}+{f_{\rm h}}^2\right)^{1/2}-{\frac
{1}{2}}f_{\rm h}, \label{h_1}
\end{equation}
where
\begin{equation}
f_{\rm h}={\frac {1}{2(1-{\rm e}^{-1/2})\kappa_0}}\left({\frac
{B_z^2}{4\pi\rho{R}H\Omega_{\rm K}^2\kappa_0}}+{\frac {\xi_{\rm
bz}B_z^2}{4\pi\rho{R^2}\Omega_{\rm K}^2}}\right),\label{f_1}
\end{equation}
and $\xi_{\rm bz}$ is defined as
\begin{equation}
{\frac {\partial B_z}{\partial R}}=-\xi_{\rm bz}(R){\frac {B_z}{R}},
\label{xi_bz}
\end{equation}
which is to be calculated as described in Section \ref{b_config}.
This fitting formula can reproduce the numerical results quite well
\citep*[][]{2002A&A...385..289C,2011ApJ...737...94C}. It reduces to
$H=c_{\rm s}/\Omega_{\rm K}$ in the absence of magnetic field. }


\subsection{Magnetic field configuration of the
NDAF}\label{b_config}

The induction equation of large-scale (poloidal) magnetic fields is
\begin{displaymath}
{\frac {\partial}{\partial t}}[R\psi(R,0)]=-v_{R}{\frac
{\partial}{\partial R}}[R\psi(R,0)]~~~~~~~~~~~~~~~~~~~~~~~~
\end{displaymath}
\begin{equation}
-{\frac {4\pi\eta}{c}}{R\over{2H}}\int\limits_{-H}^{H}
J_{\phi}(R,z_{\rm h})dz_{\rm h}, \label{indu_1}
\end{equation}
where $\psi(R,z)$ is the azimuthal magnetic potential,
$J_{\phi}(R,z_{\rm h})$ is the current density at $z=z_{\rm h}$
above/below the mid-plane of the disk \citep*[see][for the detailed
discussion]{2011ApJ...737...94C}. The magnetic field potential
$\psi(R,z)=\psi_{\rm d}(R,z)+\psi_{\infty}(R)$
\citep{1994MNRAS.267..235L}. $\psi_{\rm d}(R,z)$ is contributed by
the currents in the accretion flow, while $\psi_{\infty}(R)=B_{\rm
ext}R/2$ is the external imposed homogeneous vertical field
\citep*[see][for the details]{1994MNRAS.267..235L}. Assuming the
azimuthal current to be distributed homogeneously in $z$-direction,
we have
\begin{equation}
J_{\phi}(R,z_{\rm h})={\frac {J_{\phi}^{\rm s}(R)}{2H}},~~~~~
\label{js}
\end{equation}
where $J_{\phi}^{\rm s}(R)$ is the surface current density. The
potential $\psi_{\rm d}$ is related to $J_{\phi}^{\rm s}(R)$ with
\begin{displaymath}
\psi_{\rm d}(R,z)={1\over c}\int\limits_{R_{\rm in}}^{R_{\rm
out}}R^\prime{\rm
d}R^\prime\int\limits_{0}^{2\pi}\cos\phi^\prime{\rm
d}\phi^\prime~~~~~~~~~~~~~~~~~~~~~~~~~~
\end{displaymath}
\begin{displaymath}
\times\int\limits_{-H}^{H}
 {\frac {J_\phi(R^\prime,z_{\rm h})}{[{R^\prime}^2+R^2+(z-z_{\rm
h})^2-2R{R^\prime}\cos{\phi^\prime}]^{1/2}}}{\rm d}z_{\rm h}.
\end{displaymath}
\begin{displaymath}
={1\over {2Hc}}\int\limits_{R_{\rm in}}^{R_{\rm out}}J_\phi^{\rm
s}(R^\prime)R^\prime{\rm
d}R^\prime\int\limits_{0}^{2\pi}\cos\phi^\prime{\rm
d}\phi^\prime~~~~~~~~~~~~~~~~~~~~~~~~~~~
\end{displaymath}
\begin{equation}
\times\int\limits_{-H}^{H}
 {\frac {1}{[{R^\prime}^2+R^2+(z-z_{\rm
h})^2-2R{R^\prime}\cos{\phi^\prime}]^{1/2}}}{\rm d}z_{\rm h}.
\label{psi}
\end{equation}

Equation (\ref{indu_1}) reduces to
\begin{equation}
-{\frac {\partial }{\partial R}}[R\psi_{\rm d}(R,0)]-{\frac
{2\pi}{c}}{\frac {\alpha c_{\rm s}R}{v_R}}\mathscr{P}_{\rm
m}J_{\phi}^{\rm S}(R)=B_0{R},\label{indu_2}
\end{equation}
for steady case, where $\mathscr{P}_{\rm m}=\eta/\nu$. This
integro-differential equation can be solved numerically to derive
the $R$-dependent surface current density $J_\phi^{\rm s}(R)$ with
the specified $\mathscr{P}_{\rm m}$ if the structure of the NDAF is
known \citep*[see][for the details]{1994MNRAS.267..235L}. The
configuration of the large-scale magnetic fields can be calculated
with the derived potential $\psi$:
\begin{equation}
B_R(R,z)=-{\frac {\partial }{\partial z}}\psi(R,z),\label{b_r}
\end{equation}
and
\begin{equation}
B_z(R,z)={\frac {1}{R}}{\frac {\partial}{\partial R}}[R\psi(R,z)].
\label{b_z}
\end{equation}

\subsection{Maximal power of jets in GRBs}\label{p_bz}

Considering the horizon of a rotating black hole to be threaded by a
large scale magnetic field, the rotating energy of the hole can be
extracted to jets through the BZ process. The detailed physics is
quite complicated, which has been extensively investigated in many
previous works. For an extreme Kerr black hole, the maximal BZ power
is
\begin{equation}
P_{\rm BZ}^{\rm max}\sim {\frac {B_{\rm h}^2}{128}}R_{\rm h}^2c,
\label{p_bz_max1}
\end{equation}
where $B_{\rm h}$ is the strength of the field at the horizon, and
$R_{\rm h}=GM_{\rm bh}/c^2$ for $a=1$
\citep*[][]{1982MNRAS.198..345M,1997MNRAS.292..887G}. This provides
an upper limit on the jet power in GRBs.

\section{Results}\label{results}

In order to calculate the global structure of the NDAF as described
in Section \ref{model}, we need to specify the boundary conditions
at the outer radius $R_{\rm out}$. There are five parameters in our
model calculations: viscosity parameter $\alpha$, the mass accretion
rate $\dot{M}$, the outer radius of the accretion disk $R_{\rm
out}$, specific angular momentum of the gas in the NDAF $j$ at
$R_{\rm out}$, and the external magnetic field strength $B_{\rm
ext}$.

A typical value of $\mathscr{P}_{\rm m}=1$ is adopted in this work
\citep*[][]{parker1979,2003A&A...411..321Y,2009A&A...504..309L,2009A&A...507...19F,2009ApJ...697.1901G}.
The black hole mass $M_{\rm bh}=3M_\odot$ is adopted in all the
calculations. We use a fixed viscosity parameter: $\alpha=0.2$, in
all the calculations, which is almost independent of the field
advection, because the diffusion of magnetic field is proportional
to the viscosity for a fixed Prandtl number. $j(R_{\rm out})=j_{\rm
K}(R_{\rm out})$ is assumed at the outer radius of the NDAF, which
affects little on the dynamics of the accretion flow. The global
solution of a NDAF required to pass the sonic point smoothly is
derived by tuning $j_{\rm in}$ carefully.

The NDAF is coupled with the magnetic field, i.e., the field
configuration is determined by the NDAF structure, and the structure
of the NDAF is also affected by the large scale field. We first
calculate the global structure of the NDAF without magnetic field,
and then the magnetic field is available based on the derived NDAF
structure. The global solution of the NDAF is re-calculated based on
this derived magnetic field. The global solutions converge usually
after several iterations, because only the structure of the inner
NDAF changes significantly due to the magnetic field.

As the pseudo-Newtonian potential is used in our calculations to
simulate the general relativistic effect (see Section
\ref{stru_ndaf}), the radial velocity of the NDAF approaches
infinity at $R=R_{\rm S}$, which is unphysical. In the calculations
of the global structure of the accretion flow, we define the radius
of the black hole horizon as $R=R_{\rm h}$ where the radial velocity
$v_{R}=c$. {It is found that $R_{\rm h}\simeq 1.3-1.4~R_{\rm S}$
depending on the values of mass accretion rate $\dot{M}$ and
external field strength $B_{\rm ext}$ adopted in the calculations.}

In Figure \ref{b_conf}, we plot the large scale magnetic field
configurations of NDAFs. The field is dragged inward efficiently by
the gas in the NDAF. The structure of the NDAFs with different
values of mass accretion rate $\dot{M}$ and external field strength
$B_{\rm ext}$ is plotted in Figure \ref{vr_l_r20}. The radial
velocity of the accretion flow with different disk parameters is
quite similar. The $R$-dependent magnetic field strengths for the
NDAFs with different values of the disk parameters are plotted in
Figure \ref{b_radius}. In Figure \ref{b_radius_b}, we show how the
magnetic field advection in the NDAF varies with the outer radius of
the accretion flow. It is not surprising that the field strength at
the region close to the horizon of the the black hole increases with
the size of the disk if the values of all the other parameters are
fixed.

For given external magnetic field strength, we calculate the
structure and magnetic field of a NDAF with an outer radius $R_{\rm
out}$ when the values of parameters, $\dot{M}$, and $B_{\rm ext}$
are specified. With the derived field strength at the black hole
horizon, we estimate the maximal jet power extracted from an extreme
Kerr black hole with Equation (\ref{p_bz_max1}). The maximal
Blandford-Znajek jet power as functions of the disk size is plotted
in Figure \ref{p_bz_max} with different values of $\dot{M}$ and
$B_{\rm ext}$. In Figure \ref{vr_b_r20}, we show how the structure
of the NDAFs changes with the mass accretion rate $\dot{M}$ for
given external magnetic field strength $B_{\rm ext}$. We find that
the radial velocity of the accretion flow decreases near the black
hole horizon if $\dot{M}$ is low, which implies the accretion flow
is dynamically controlled/arrested by the magnetic field. The
condition for arrested accretion flows is plotted in Figure
\ref{mdot_crit}.


 \figurenum{1}
\centerline{\includegraphics[angle=0,width=7.5cm]{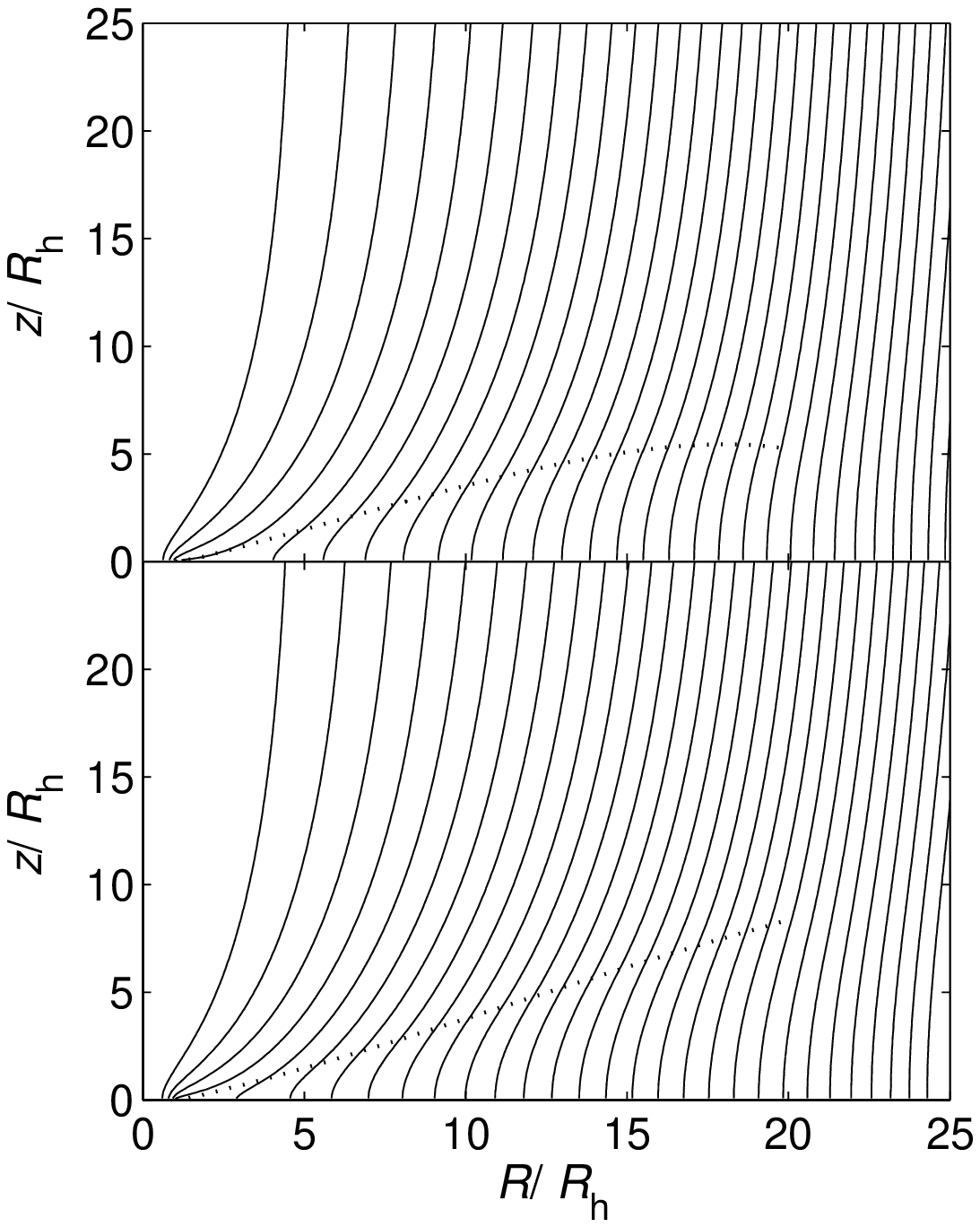}}
\figcaption{The large-scale poloidal magnetic field configurations
of NDAFs. The outer radius of the disk is assumed to be $R_{\rm
out}=20R_{\rm h}$. The dotted line is the scale height of the NDAF.
The magnetic field configuration plotted in the upper panel is
calculated with $\dot{M}=0.01 M_\odot~s^{-1}$ and $B_{\rm
ext}=10^{13}$~Gauss, while the lower panel is for $\dot{M}=1
M_\odot~s^{-1}$ and $B_{\rm ext}=10^{14}$~Gauss. Every magnetic
field line corresponds to a certain value of the stream function
$R\psi(R,z)$ with the lowest $[R\psi(R,z)]_{\rm min}=0.1R_{\rm
out}\psi_{\infty}(R_{\rm out},0)$ increasing by $\Delta
R\psi(R,z)=0.1R_{\rm out}\psi_{\infty}(R_{\rm out,0})$ per line,
where $\psi_{\infty}(R_{\rm out})=B_0R_{\rm out}/2$.
\label{b_conf}}\centerline{}


 \figurenum{2}
\centerline{\includegraphics[angle=0,width=7.5cm]{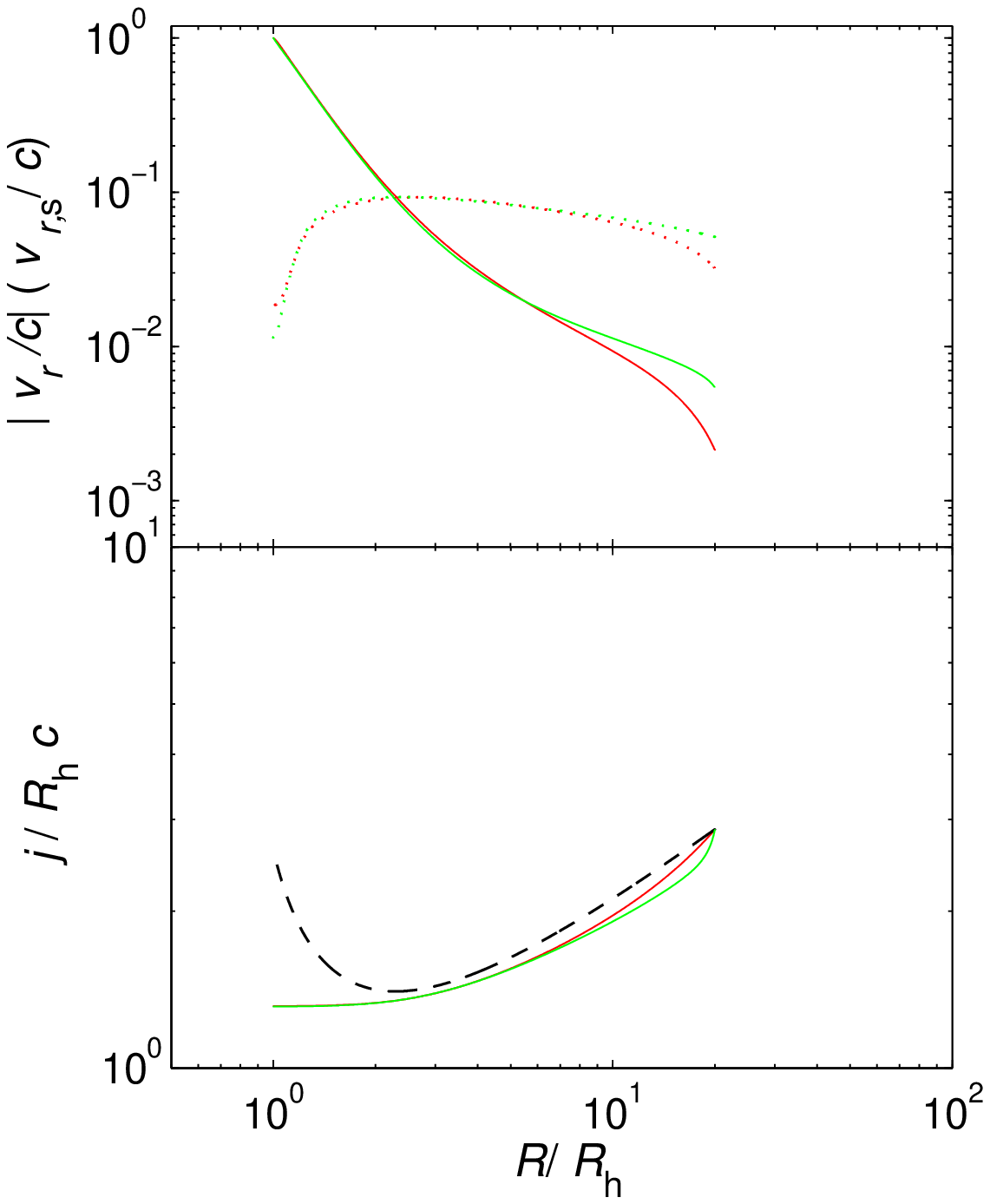}}
\figcaption{The structure of NDAFs with different values of disk
parameters. The red lines indicate the results calculated with
$\dot{M}=0.01 M_\odot~s^{-1}$ and $B_{\rm ext}=10^{13}$~Gauss, while
the green lines are for the results with $\dot{M}=1 M_\odot~s^{-1}$
and $B_{\rm ext}=10^{14}$~Gauss. The radial velocity (solid lines)
and sound speed (dotted lines) are plotted in the upper panel. In
the lower panel, the specific angular momenta as functions of radius
are plotted. The black dashed line represents the Keplerian specific
angular momentum.
 \label{vr_l_r20}}\centerline{}


 \figurenum{3}
\centerline{\includegraphics[angle=0,width=7.5cm]{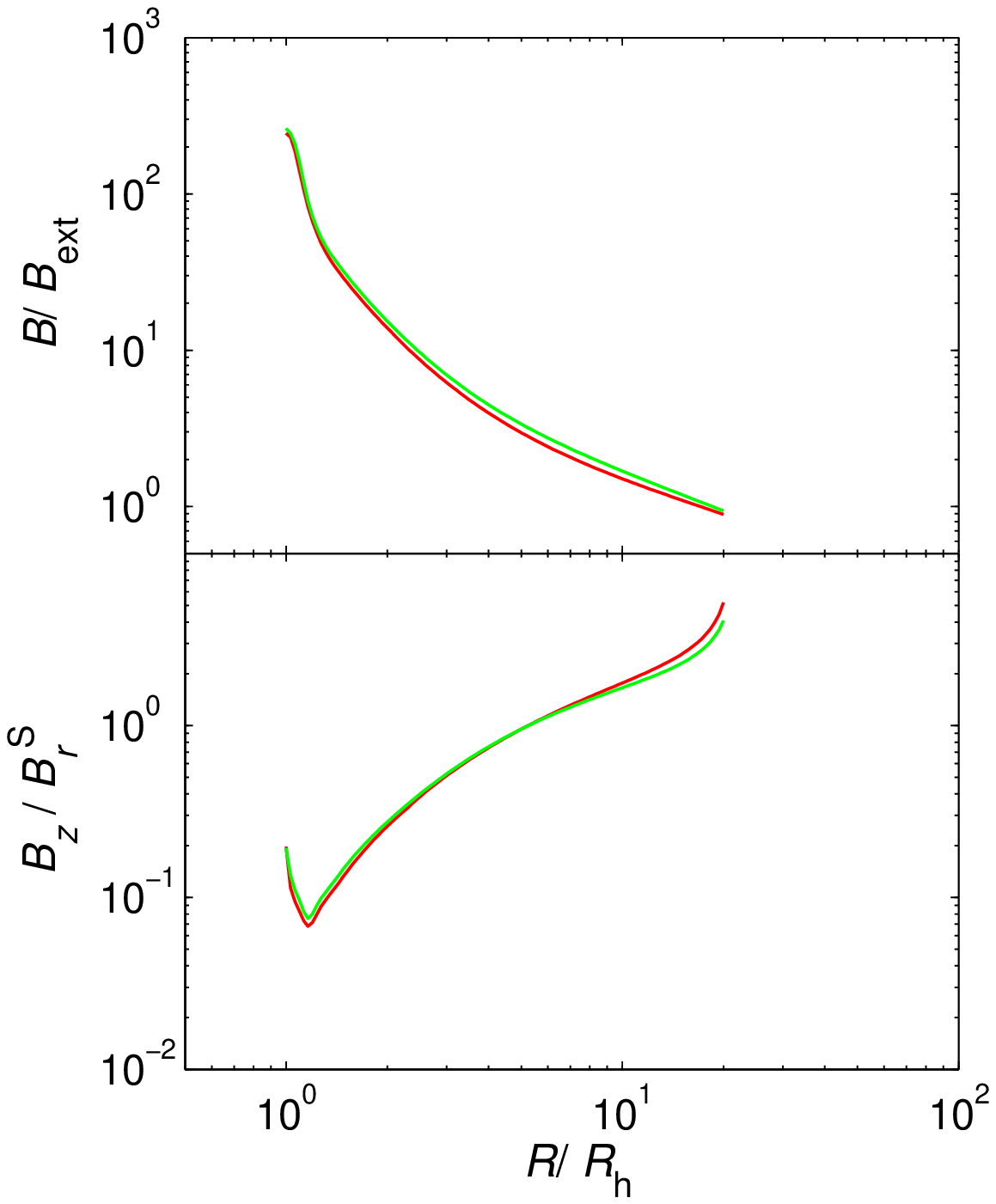}}
\figcaption{The magnetic field strengths as functions of radius for
the NDAFs with different values of model parameters are plotted in
the upper panel. The inclinations of the field lines at the disk
surface are plotted in the lower panel. The red line indicates the
result calculated with $\dot{M}=0.01 M_\odot~s^{-1}$ and $B_{\rm
ext}=10^{13}$~Gauss, while the green line is for the result with
$\dot{M}=1 M_\odot~s^{-1}$ and $B_{\rm ext}=10^{14}$~Gauss.
\label{b_radius} }\centerline{}


\figurenum{4}
\centerline{\includegraphics[angle=0,width=7.5cm]{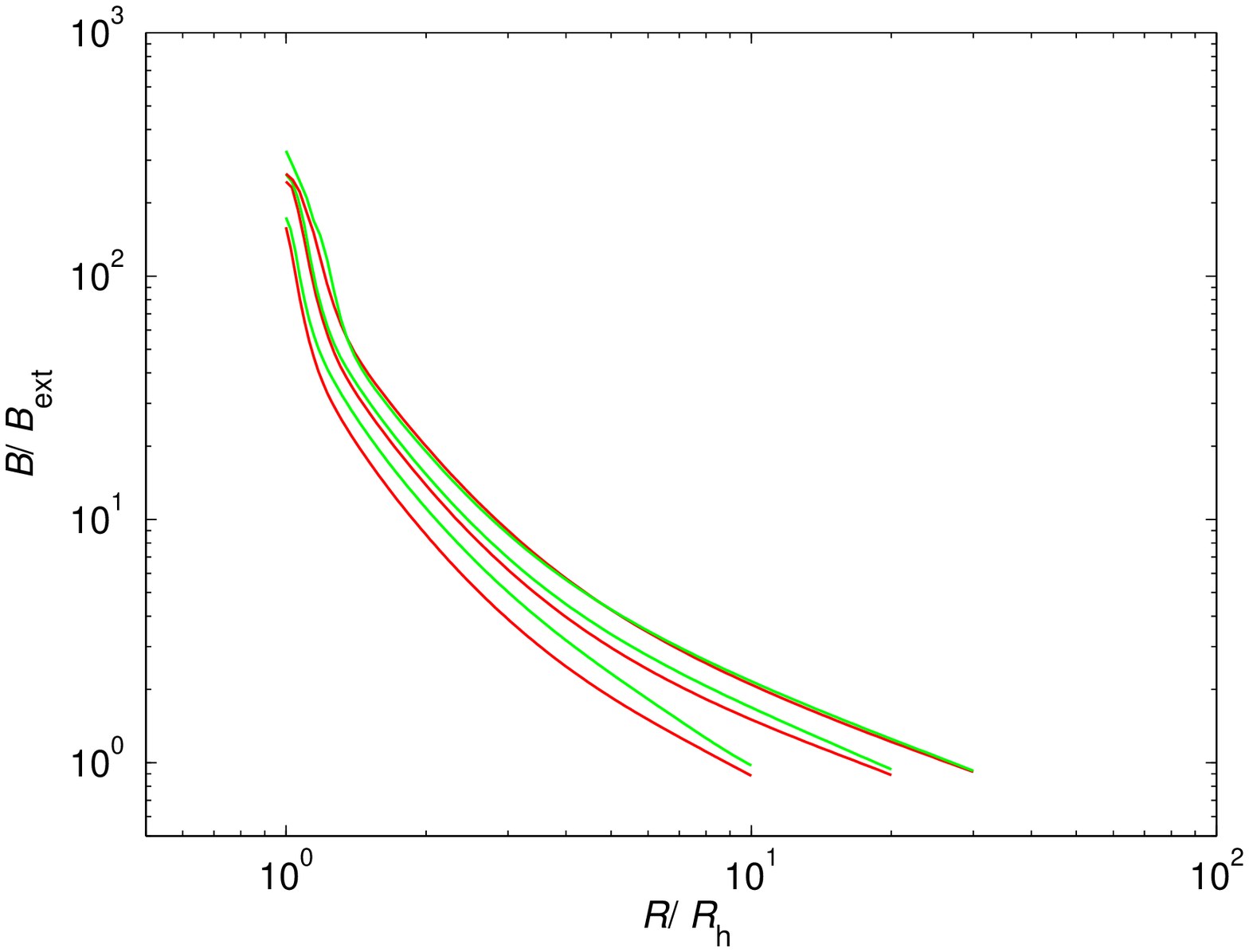}}
\figcaption{The magnetic field strengths as functions of radius for
the NDAFs with different outer radii of NDAFs ($R_{\rm out}=10R_{\rm
h}$, $20R_{\rm h}$, and $30R_{\rm h}$). The red lines indicate the
results calculated with $\dot{M}=0.01 M_\odot~s^{-1}$ and $B_{\rm
ext}=10^{13}$~Gauss, while the green lines are for the results with
$\dot{M}=1 M_\odot~s^{-1}$ and $B_{\rm ext}=10^{14}$~Gauss.
 \label{b_radius_b}}\centerline{}


\figurenum{5}
\centerline{\includegraphics[angle=0,width=7.5cm]{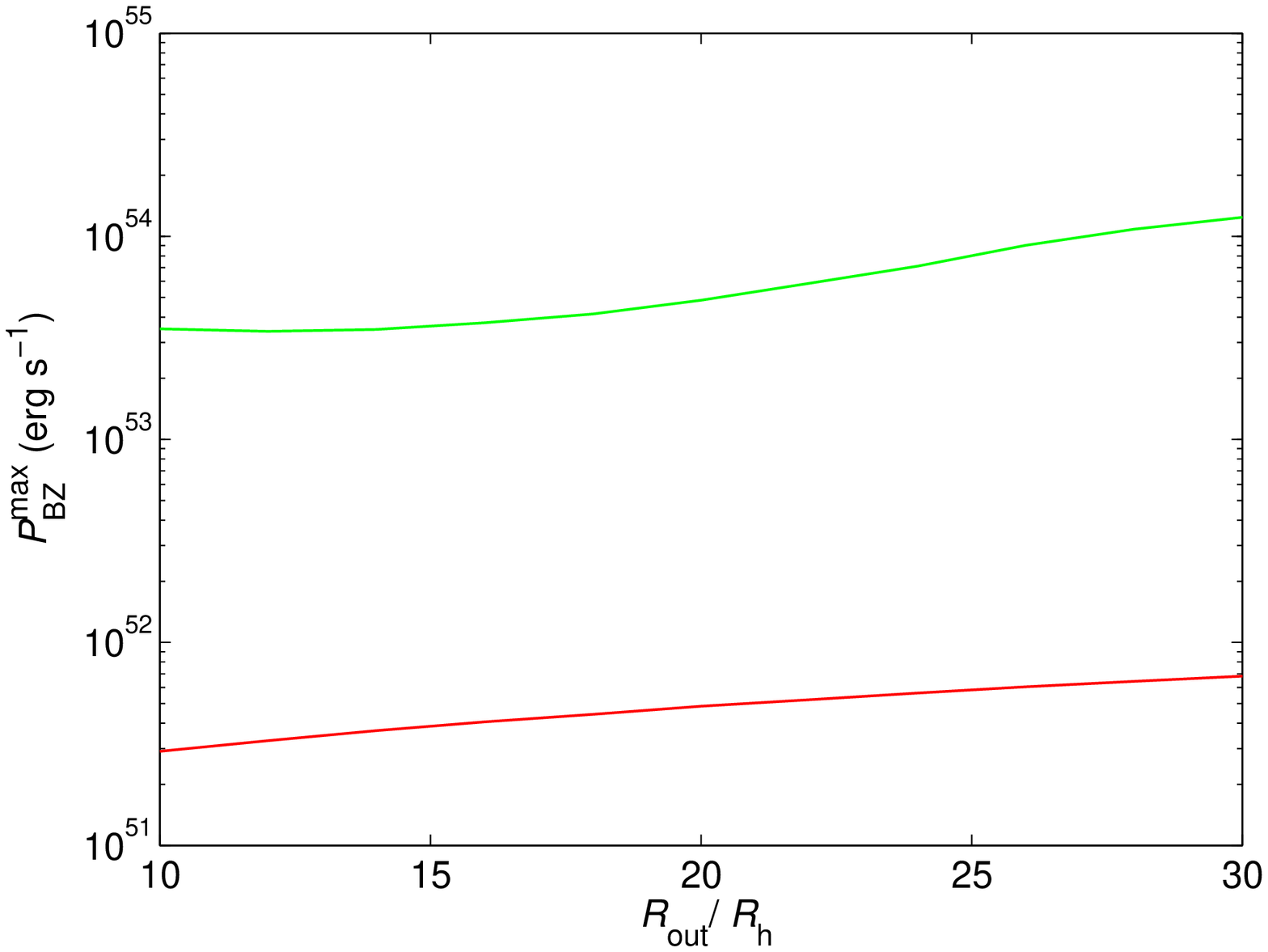}}
\figcaption{The maximal jet power can be extracted
 from extreme Kerr black holes surrounded by NDAFs. The red lines indicate
the results calculated with $\dot{M}=0.01 M_\odot~s^{-1}$ and
$B_{\rm ext}=10^{13}$~Gauss, while the green lines are for the
results with $\dot{M}=1 M_\odot~s^{-1}$ and $B_{\rm
ext}=10^{14}$~Gauss.
 \label{p_bz_max}}\centerline{}


\figurenum{6}
\centerline{\includegraphics[angle=0,width=7.5cm]{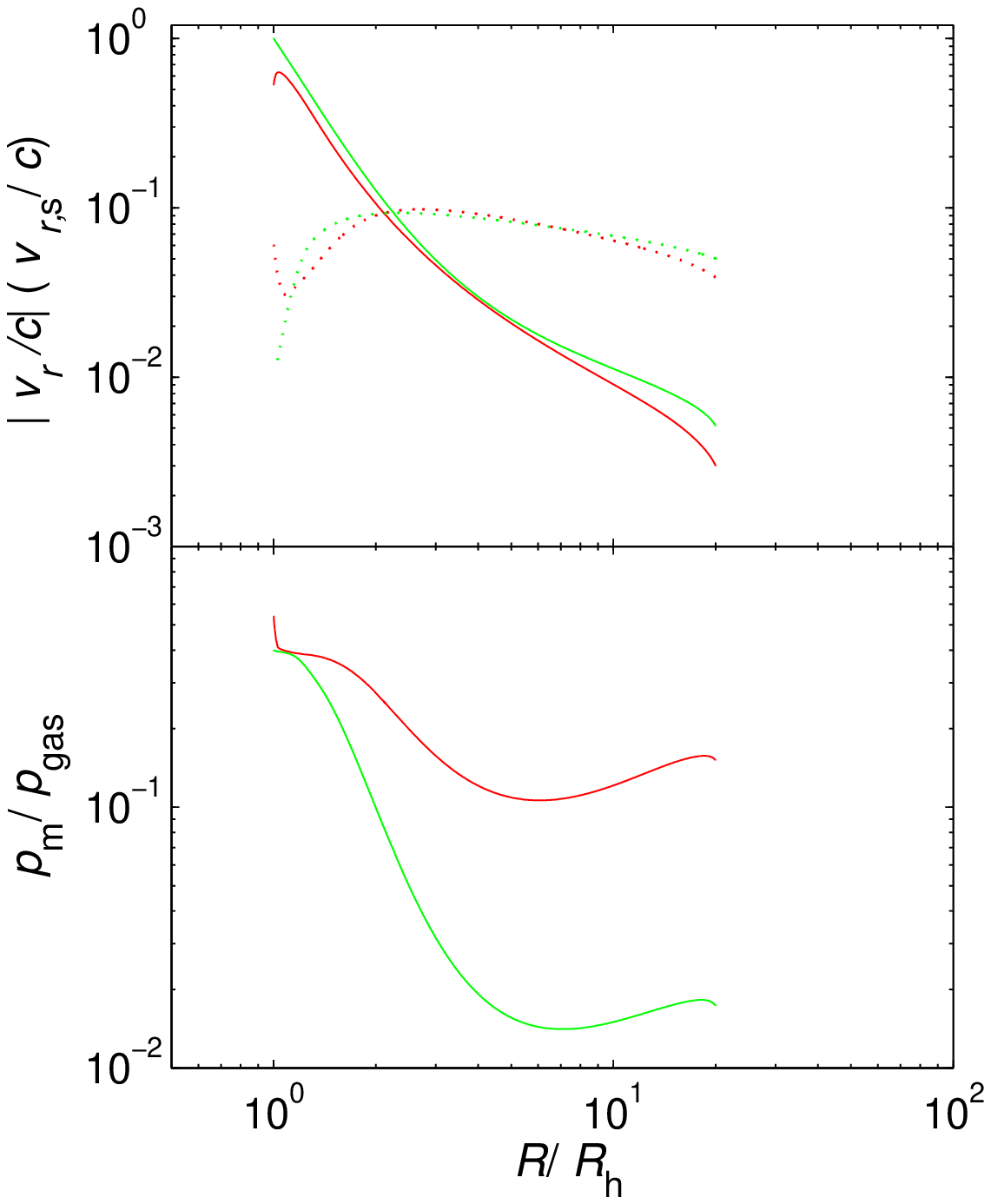}}
\figcaption{The structure and magnetic field strength
 of NDAFs. The strength of the external magnetic field $B_{\rm
 ext}=10^{14}$~Gauss is adopted in the calculations.
The red lines indicate the results calculated with $\dot{M}=0.1
M_\odot~s^{-1}$, while the green lines are for the results
calculated with $\dot{M}=1 M_\odot~s^{-1}$. The radial velocity
(solid lines) and sound speed (dotted lines) are plotted in the
upper panel. In the lower panel, we plot the ratio of $p_{\rm
m}/p_{\rm gas}$ as a function of radius. \label{vr_b_r20}
}\centerline{}


 \figurenum{7}
\centerline{\includegraphics[angle=0,width=7.5cm]{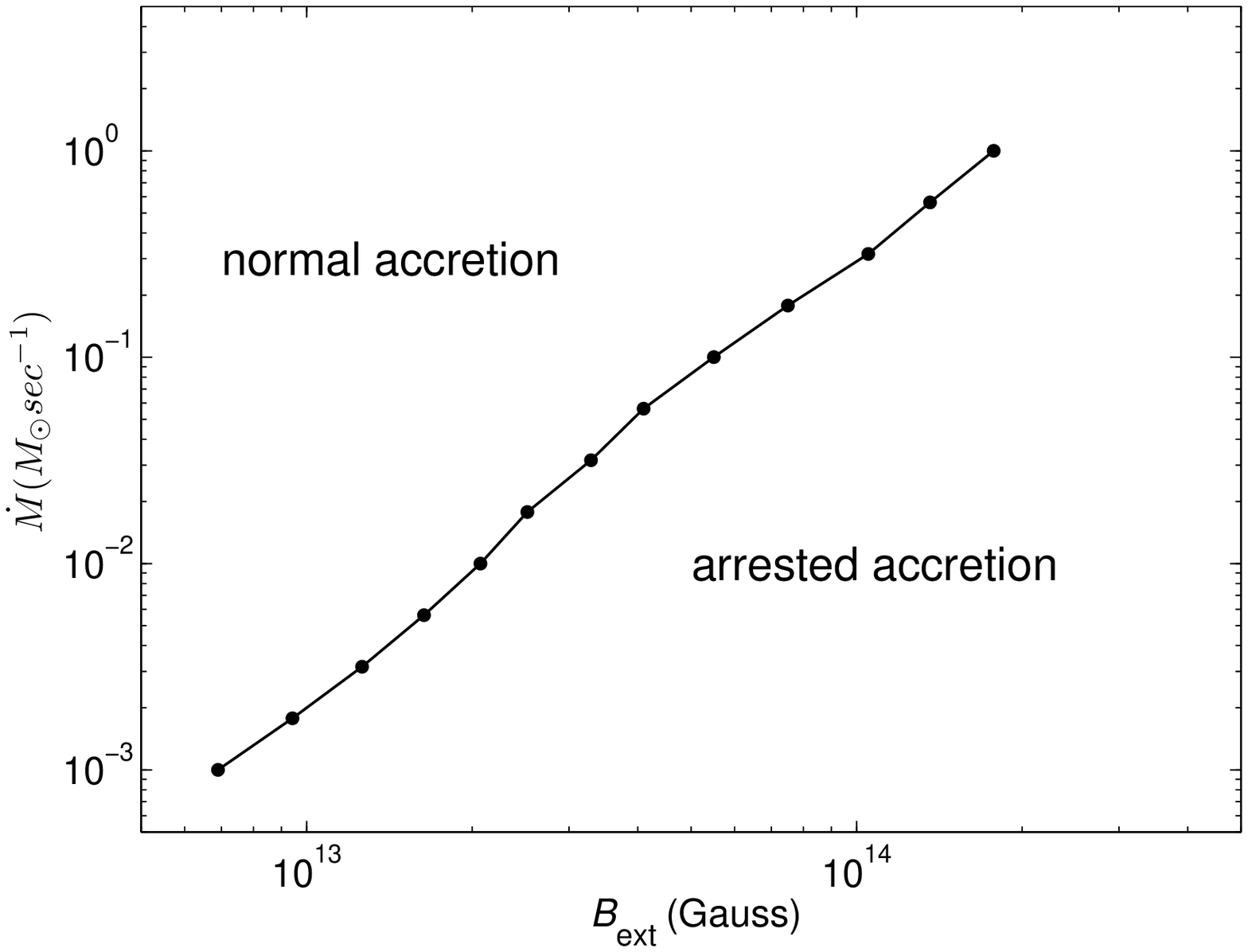}}
\figcaption{The critical mass accretion rate
 $\dot{M}$ for magnetically arrested accretion as a function of the external field strength $B_{\rm
 ext}$. \label{mdot_crit}}\centerline{}

\section{Discussion}\label{discussion}

The external field can be dragged inward by the NDAF efficiently
(see Figure \ref{b_conf}). {The magnetic field can hardly be
advected by a geometrically thin standard accretion disks for a
typical value of $\mathscr{P}_{\rm m}\sim 1$
\citep{1994MNRAS.267..235L}. The relative disk thickness $H/R$ is
large in the outer region of the NDAF (see Figure \ref{b_conf}),
which implies the field advection is more efficient in NDAF than the
thin disk case considered in \citet{1994MNRAS.267..235L}. The disk
can be geometrically thin in the inner region of the NDAF, however,
the gas falls almost freely onto the black hole horizon within the
inner stable circular orbit of the black hole. Its radial velocity
approaches the light speed at the black hole horizon, and therefore
the external field is also dragged efficiently in the inner region
of the NDAF towards the black hole. This is different from the
calculations given in \citet{1994MNRAS.267..235L}, in which the
transonic characteristics of the accretion flow near the black hole
have not been included. In their work, the gas in the standard thin
disk is assumed to rotate at Keplerian velocity $v_{\rm K}$, and its
radial velocity of the disk is approximated as
$v_R=-3\nu/2R=-(3/2)(H/R)^2\alpha v_{\rm K}$, which is
underestimated in the inner region of the disk. }

For a given external magnetic field, the advection of the field in
the NDAF can be calculated. We find that the field can be amplified
more significantly for a NDAF with a larger size (a larger outer
radius $R_{\rm out}$), due to the external field in a large area be
captured by the accretion flow \citep*[see Figure
\ref{b_radius_b},][]{2003PASJ...55L..69N}. The strength of the
advected field at the black hole horizon is derived in our
calculations, and then the maximal jet power is estimated. In
principle, the calculations for the NDAF surrounding a rotating
black hole should be carried out in the general relativistic frame
\citep{2013arXiv1306.0655X}, which is too complicated for our
present investigation. The calculations based on the
pseudo-Newtonian potential in this work show the main feature of the
general relativistic accretion flow, i.e., the radial velocity
approaches the light speed at the black hole horizon. We believe the
estimates of the BZ jet power in this work based on the
pseudo-Newtonian potential should be reliable at order of magnitude.
In Figure \ref{p_bz_max}, we show the maximal Blandford-Znajek jet
power as functions of the disk size with different values of
$\dot{M}$ and $B_{\rm ext}$. Our results show that the maximal BZ
jet power can be as $\sim 10^{53}-10^{54}$~erg~s$^{-1}$ for an
extreme Kerr black hole with 3$M_\odot$, if an external magnetic
field with $10^{14}$~Gauss is advected by the NDAF accreting at
$\dot{M}=1M_\odot~{\rm s}^{-1}$, which is consistent with the
typical field strength of a tidally disrupted magnetar
\citep*[][]{2012ApJ...760...63L}. {The tidal radius of a star
disrupted by a black hole can be estimated by
\begin{equation}
R_{\rm T}\simeq 7.2 \left({\frac {M_{\rm bh}}{M_*}}\right)^{1/3}R_*,
\label{r_t}
\end{equation}
where $M_*$ and $R_*$ are the mass and radius of the star
respectively \citep*[][]{1988Natur.333..523R}. For a typical
magnetar with $M_*=1.4 M_\odot$ and $R_*=10$~km captured by an
extreme Kerr black hole with $M_{\rm bh}=3M_\odot$, the tidal radius
$R_{\rm T}\simeq 21 R_{\rm h}$, which should be comparable with the
outer radius of the NDAF.  }

{An episodic soft gamma-ray tail is preferred to show in short-type
GRBs, which are believed to be produced by mergers of compact stars
\citep*[][]{Hu2014}.} The steady configuration of the field dragged
by the NDAF is achieved when the advection is balanced with magnetic
diffusion \citep*[see][for the detailed
discussion]{1994MNRAS.267..235L}. We find that the accretion flow is
arrested by the magnetic field close to the black hole horizon if
the mass accretion rate $\dot{M}$ is low (see Figure
\ref{vr_b_r20}). There is a critical mass accretion rate for a given
external magnetic field, below which the accretion flow is arrested
(see Figure \ref{mdot_crit}). This implies that the accretion may be
oscillating, because the balance between advection and diffusion is
destroyed. When the accretion flow is arrested by the magnetic
field, the accretion stops and the magnetic field decays due to
magnetic diffusion. The accretion re-starts when the field becomes
sufficiently weak, and the accretion flow may be arrested again if
the advected field is strong enough. This will lead to episodic
accretion, which is suggested to explain the observed oscillating
soft extend emission in short-type GRBs
\citep*[][]{2006MNRAS.370L..61P,2012ApJ...760...63L}.

As discussed above, the time interval of the episode accretion
should be comparable with the magnetic field diffusion timescale. We
can estimate the diffusion timescale of the field as
\citep*{1994MNRAS.267..235L}
\begin{displaymath}
t_\eta\sim{\frac {R^2}{\eta}}{\frac {H}{R}}{\frac {B_z}{B_r^{\rm
S}}}={\frac {1}{\alpha\mathscr{P}_{\rm m}\Omega_{\rm K}}}{\frac
{B_z}{B_r^{\rm S}}}\left({\frac
{H}{R}}\right)^{-1}~~~~~~~~~~~~~~~~~~~~~~~~~~~~~
\end{displaymath}
\begin{equation}
=4.92\times 10^{-6}{\frac {M}{M_\odot}}{\frac {B_z}{B_r^{\rm
S}}}\left({\frac {R}{R_{\rm h}}} \right)^{3/2}\left({\frac
{H}{R}}\right)^{-1}\alpha^{-1}\mathscr{P}_{\rm m}^{-1}~{\rm s}.
 \label{t_eta}
\end{equation}
The poloidal magnetic field of the NDAF is maintained by the
azimuthal currents in the whole disk, which is a global problem, and
therefore we adopt typical values of the quantities at the outer
radius of the NDAF in estimating the diffusion timescale of the
field threading the disk. If we adopt $H/R=0.2$, and $B_z/B_r^{\rm
S}=5$ at $R_{\rm out}$ (see Figures \ref{b_conf} and
\ref{b_radius}), the diffusion timescale $t_\eta\sim 0.2$ seconds,
for the typical values of the parameters, $\mathscr{P}_{\rm m}=1$,
$\alpha=0.2$, and $R=20R_{\rm h}$. The size of the disk may expand
with time due to the angular momentum transfer in the disk
\citep*[see Figure 5.1 in][]{2002apa..book.....F}, which implies
that the radius of the arrested NDAF at late stage may be larger
than the tidal radius $R_{\rm T}$. Thus, the interval timescale of
the episodic accretion could be longer than the estimate, which is
roughly consistent with the observed oscillation in the soft extend
emission of short-type GRBs \citep*[see Figure 2 in][for the
observed light curves of some typical GRBs]{2012ApJ...760...63L}.

\acknowledgments We thank the referee for his/her helpful
comments/suggestions, and Dong Lai for discussion. This work is
supported by the NSFC (grants 11173043, 11121062, 11233006, and
11025313), the CAS/SAFEA International Partnership Program for
Creative Research Teams (KJCX2-YW-T23), the Strategic Priority
Research Program of the Chinese Academy of Sciences (XDB09000000).
E. W. Liang acknowledges support by the National Basic Research
Program (973 Programme) of China (Grant 2014CB845800) and the
Guangxi Science Foundation (2013GXNSFFA019001).

{}

\end{document}